**Electric Field Assisted Nanostructuring of a Mott Insulator**

By *Vincent Dubost\*, Tristan. Cren, Cristian Vaju, Laurent Cario, Benoit Corraze, Etienne Janod, François Debontridder, Dimitri Roditchev*

[*]V. Dubost, Dr. T. Cren, Dr. F. Debontridder, Dr. D. Roditchev,
    Institut des Nanosciences de Paris (INSP), CNRS UMR 75-88, Université Paris 6 (UPMC) 140 rue de Lourmel, 75015 Paris, France
E-mail : vincent.dubost@insp.jussieu.fr

Dr. C Vaju, Dr. L. Cario, Dr. B. Corraze, Dr. E. Janod
    Institut des Matériaux Jean Rouxel (IMN), Université de Nantes, CNRS, 2 rue de la Houssinière, 44322 Nantes Cedex 3, France
[**]    The authors thank Julie Martial at IMN for her help in samples elaboration, and Jean-Charles Ricquier for his work on the figures. This work was supported by a Young Researcher grant ANR-05-JCJC -0123-01 from the french Agence Nationale de la Recherche (to L. C., B.C and E. J.)

**Abstract**

We report the first experimental evidence for a strong electromechanical coupling in the Mott insulator $GaTa_4Se_8$ allowing a highly reproducible nano-writing with a Scanning Tunneling Microscope (STM). The local electric field across the STM junction is observed to have a threshold value above which the clean (100) surface of $GaTa_4Se_8$ becomes mechanically instable: At voltage biases V > 1.1V the surface suddenly inflates and comes in contact with the STM tip, resulting in nanometer size craters. The formed pattern can be indestructibly "read" by STM at lower voltage bias, thus allowing a 5 Tdots/inch$^2$ dense writing/reading at room temperature. The discovery of the electromechanical coupling in $GaTa_4Se_8$ might give new clues in the understanding of the Electric Pulse Induced Resistive Switching recently observed in this stoechiometric Mott insulator.

# 1. Introduction

Electric Pulse Induced Resistive Switching (EPIRS), which accounts for a non volatile change of resistance upon application of an electric pulse, shows a current burst of activity owing to possible applications in non-volatile memory devices.[1] So far, three mechanisms i.e. Joule heating, interfacial electronic injection, and ionic electro-migration were essentially evoked to explain the EPIRS phenomena[1] encountered in materials such as NiO, [2] manganites,[3] and $SrTiO_3$,[4] respectively. However, recently discovered EPIRS phenomenon in the non-centrosymmetric fragile Mott Insulator $GaTa_4Se_8$ [5] is not described by any of these mechanisms. Due to its peculiar crystalline and electronic structures, this clustered lacunar spinel compound lies on the verge of the Metal-Insulator transition which makes it very sensitive to external parameters. Thus, $GaTa_4Se_8$ undergoes an Insulator-to-Metal and even superconducting transition under pressure.[6] More unexpectedly, electric pulses[5] induce in this compound an electronic phase separation on a nanometer scale provoking an insulator to granular metal/superconductor transition.[7] While both pressure and electric field provoke an Insulator Metal transition, it is not clear if the resulting metallic states are identical at the microscopic scale. In this Letter we report a STM experiment performed on the clean (100) surface of $GaTa_4Se_8$ that gives an experimental evidence of a strong coupling, down to the nanoscale, between the electrical and mechanical properties in this material. We show that the surface of a $GaTa_4Se_8$ crystal becomes mechanically unstable upon the application of a local electric field across the STM junction, above a certain threshold. This strong electro-mechanical effect enables a nanoscale etching of the surface which was used to create nanostructured patterns stable at room temperature, their density could be as high as 5 Tdots/inch$^2$. Moreover our discovery supports, in a complementary manner, a new EPIRS mechanism based on the Mott transition proposed for $GaTa_4Se_8$ in Vaju et al..[5]

Our previous preliminary STM study had revealed that pristine $GaTa_4Se_8$ crystals were hardly measurable, while crystals which had undergone an EPIRS prior to the measurement were much more suited for STM studies.[5] In that respect, in the present report we have focused on so-called weakly transited $GaTa_4Se_8$ crystals that were prepared as follows. First of all, four electric contacts were glued onto a small single pristine crystal (typical crystal size ~100-300microns) of $GaTa_4Se_8$. Then, at T= 77 K, a series of electric pulses of increasing intensities were applied through the current leads while the conductance state was controlled via the voltage leads.[7] At this temperature the fully transited crystals show a resistance change over several orders of magnitude[5] upon switching. However in the present study, in order to obtain weakly transited crystals, the series of pulses was stopped when the resistance of the sample after the pulse was found to be roughly half of its value in the pristine insulating state. The non volatile character of the transition was confirmed in both four probe and two probe configurations, thus indicating its bulk nature. This weakly transited state remained stable at room temperature for several weeks. All the STM results described hereafter were obtained on such weakly transited crystals.

The weakly transited crystals were carefully oriented and glued on our STM sample holder and introduced into the Ultra High Vacuum (UHV) STM chamber (base pressure $7.10^{-11}$ mbar). Then, they were cleaved *in-situ* along the (100) planes using a hardened-steel blade. Electrochemically etched tungsten STM tips were thermally annealed under UHV up to 1300-1400K by direct current heating before use in order to get read of oxide at the tip apex. The large scale topographic STM images acquired on such *in-situ* cleaved surfaces display large terraces, several hundred of nanometers wide **(Figure 1a)**. These terraces are notched by 0.5nm deep trenches of various widths, all aligned in the same direction. The direction of the trenches can be related to the crystal structure of $GaTa_4Se_8$. Indeed, the crystal structure of the material[6,8] corresponds to a rocksalt-type packing of $Ta_4Se_{16}$ clusters and $GaSe_4$ tetrahedra, sharing Se atoms. Hence, a cleavage parallel to (100) likely breaks Ta-Se bonds, exposing a slab where $Ta_4Se_{16}$ clusters and $GaSe_4$ tetrahedra alternate. The thickness of such a slab is half the unit cell parameter, i.e. 0.5 nm, that corresponds to the depth of the observed trenches. The surface space group is p2mm and the direction of the trenches corresponds to the [1-1] direction, parallel to the rows of $GaSe_4$ tetrahedra and $Ta_4Se_{16}$ clusters **(Figure 1b).** This kind of pattern, as well as its close relation to the crystal structure is a strong evidence of an important surface relaxation[1]. A zoom on an atomically flat terrace is shown **Figure 2a**. Despite its disordered appearance, the surface has a very small roughness of typically 100 pm. The self-correlation analysis allows estimating a characteristic lateral size for the observed bumps to roughly 3 nm. Despite numerous attempts, atomic resolution imaging has not been achieved on this surface[2]**;** spatial inhomogeneities have been observed everywhere on the surface; their overall pattern is reproducible from one sample to another.

The tunnelling junction was characterized using current versus distance spectroscopy I(Z). As the tip is retracted, the I(Z) spectra display an exponential decay. Assuming a standard expression for the tunnelling current $I(Z) = \exp(-2\kappa Z)$, where $\kappa = \sqrt{2m^* W}/\hbar$ and $m^* = m_e$, one obtains a high work function W = 5.6 +/- **0.2** eV, evidencing a clean vacuum tunneling junction.

**Results and discussion**

The influence of voltage pulses was studied in the following manner. First, the tip position was regulated with a typical current set-point I = 0.1 nA for an applied voltage across the junction $V_{bias}$ = 0.7 V. Then, the tip position was fixed by switching off the STM feedback loop, and a rectangular bias pulse was applied for the desired time. At the end of the pulse, the bias voltage was switched back to its initial value, and the STM feedback loop was switched on. While such pulses did not produce any effect at low biases, for bias pulses higher than 1.1V we systematically observed a sudden saturation of the STM current preamplifier. This is a clear signature of the appearance of a conducting link between the STM tip and the sample surface[3]. In order to evaluate the effect of the pulses on the surface, we performed constant current STM images of the perturbed region with a moderate $V_{bias}$~0.7V at which the tunnelling junction always showed a conventional behaviour. The STM images revealed that the pulses created nanometre scale craters with a typical depth of 1 to 3 nm, as measured with respect to the unmodified surface (see Fig 3a). Their inner structure displayed a pattern with a well reproducible shape that we attributed to the footprint of the tip. This is quite similar to what Carrasco et al.[9] found in nano-indentation experiments on Au. The periphery of the crater is roughly circular. Such a peculiar crater shape is consistent with the formation of a tip-surface nano-contact upon bias pulses.

To analyze the effect of voltage pulses on the crater size we applied pulses on a grid varying both the pulse voltage and duration, as shown in Fig 3. The craters in the test lines (red arrows in Fig. **3a**) were created using 100 μs, 500 μs, 1 ms, 5 ms, 20 ms, 50 ms and 100 ms pulse duration (from bottom to the top) and pulse voltage varying from +1.15V to +1.9 V by steps of 0.05 V (from left to right). On the other hand, all the craters of the reference grid lines (blue arrows in Fig. **3a)** were made with identical 100 ms pulses of 1.6V. The reference lines were thus used to check the reproducibility of the crater creation[4].The size distribution of the reference craters is surprisingly narrow, as the histogram in Fig.**3b** shows. One should note the very low rate of unsuccessful attempts i.e. missing crater (in Fig. 3a), of the order of 2 %. The crater radius distribution nicely follows a Gaussian statistics, with an average radius of 5.5 nm and a radius variation of about 10 %**.** Going back to the test lines, it comes directly from Fig.**3a** that the pulse amplitude is a relevant parameter in the crater creation, while the pulse duration does not seem important, at least in the studied range. In Fig. **3c** we have plotted the variation of the crater radius as a function of the pulse amplitude. The crater radius increases linearly with voltage pulse; an extrapolation to zero gives a voltage threshold of 1.1 V for the crater creation. We also checked the influence of the set point (i.e. of the tip-to-sample distance) on the crater formation. To do so, before applying the voltage pulse, we moved the tip towards the sample by a controlled distance ranging from 0 to 2 Angstroms (keeping the feed back loop of the STM open). Within the studied range, we did not observe any significant dependence of the crater size and shape on the tip-sample distance. We argue that in the Mott insulator the screening is poor, and the electric field of the tip spreads well inside the material and consequently, it is not so sensitive to variations of a few Angstroms of the tip-sample distance. Finally, we studied the influence of the sign of the voltage pulse but we did not find any significant difference between opposite voltage pulses.

Using the process described above we were able to routinely create thousands of reproducible 10-15 nm wide craters at room temperature. In figure 4 we give an example of such a 10 nm resolution "writing". Remarkably, it was impossible to "etch" an entire region with such voltage pulses, i.e. to create continuous "cratered" regions. In fact, the probability to create a second crater in the vicinity of an existing one is very low for the separation distances smaller than 10nm. The origin of this effect is still unknown. Hence, with 100ms@1,6V pulses the resolving power of such a nano-writing is 10 nm, which corresponds to a density of about 5 TeraDots/inch$^2$. This density can be further increased by optimizing the pulse voltages, the smallest reproducible craters were created with 1ms@1.2V pulses, giving a typical radius of 1nm. Once created, these patterns are remarkably stable, and appear unmodified at room temperature for at least one week. Continuous scanning at voltages below the threshold value (1.1V) does not affect the created structures.

Let us address now the possible mechanisms responsible for the surface nanostructuration. In fact, the nanoscale modification of surface with a STM tip has been extensively studied in the past. Besides indentations on soft metals[9] or material deposition from the tips, different models for nanostructuring with STM have been proposed: mechanical contact, field evaporation, Joule heating, electrochemical reactions. Electrochemical reactions induced by the electric field of the tip as on $SrRuO_3$, Cuprates, Gold, $WSe_2$, manganites[10] are possible only in air due to the presence of a thin water film on the surface. In our case however, the crystals were cleaved under UHV and consequently, the effect should be intrinsic to the clean surface of $GaTa_4Se_8$. The fact that the crater radius does not depend on the pulse duration while it varies linearly with the voltage pulse seems to exclude thermal effects as a possible origin. Field evaporation and local fusion[11] have also been proposed as possible origins for nanostructuring on phase change materials; nevertheless, the voltage threshold for the field evaporation effect on clean surfaces is of the order of the binding energy which corresponds to several volts. On silver chalcogenides ($Ag_2S$, $Ag_2Se$), the application of an electric field of the same order of magnitude as in our case, in STM as well as in nano-bridge geometries, can generate a hole or a neck.[12] This mechanism involves the displacement of $Ag^+$ ions and is odd regarding to the sign of the electric field which is not consistent with our results. Therefore none of the previously proposed mechanisms for nanostructuring the surface with a STM tip seems to apply in the case of $GaTa_4Se_8$. Conversely, all our observations for this compound are consistent with a mechanical instability of the surface upon the application of a local electric field. This surface inflation below the tip suggests the existence of an electromechanical coupling in $GaTa_4Se_8$. As the structure of $GaTa_4Se_8$ is non centrosymmetric and compatible with piezoelectricity, a straightforward explanation of this phenomenon could be a piezoelectric response of the material. However, piezoelectricity is a linear, and hence odd, relation between the strain and the electric field, which is not consistent with our experimental findings. At this step it is therefore not yet possible to conclude concerning the nature of the observed electromechanical coupling in $GaTa_4Se_8$. However this finding might provide a new insight in the previously observed resistive switching. Indeed, $GaTa_4Se_8$ exhibits both a pressure and an electric pulse induced insulator-metal transition. An electromechanical coupling which would convert the electric field in a kind of "pressure effect" might therefore be the missing link between the application of an electric pulse and the modification of the resistance state of $GaTa_4Se_8$.[5] In that respect the electro-mechanical coupling evidenced in our experiments supports a new mechanism of EPIRS phenomenon in $GaTa_4Se_8$ i.e. an electric pulse breakdown of the fragile Mott insulating state of this compound.

**Conclusions**

In conclusion, we have demonstrated that the UHV cleaved (100) surface of $GaTa_4Se_8$ becomes mechanically unstable upon the application of a local electric field. As a result, short voltage pulses above the threshold bias $V_{thershold}$ ~1.1 Volts produce sharp holes at the surface. This process, which depends neither on sign nor on pulses duration, can be used for nanostructuring the surface, allowing a pattern density up to 5 TeraDot per inch$^2$ stable at room temperature. This effect cannot be related to the mechanisms for nanoscale modifications of the surface reported until now. The most probable origin is linked to the presence of a strong electro-mechanical coupling in $GaTa_4Se_8$. In the near future, conducting tip Atomic Force Microscopy experiments will allow following the bulging of the surface in a more direct way. Finally, the observed electro-mechanical coupling might give new clues to understanding the striking Electric Pulse Induced Resistive Switching recently observed in this compound.

**Experimental**

[1] *Surface caracterization*: At this stage of study, the driving force of this surface relaxation remains unknown; however, when cleaved in air the samples do not show such a relaxation.

[2] *STM Resolution* : Note that our STM set-up has a resolution better than 1pm as determined by atomic resolution imaging on noble metal surfaces like Silver (111) obtained at room temperature

[3] *Caracterization of the effect :* Such a puzzling effect was observed uniquely on AEP $GaTa_4Se_8$ crystals. It was never observed in our STM set-up when studying Ag(111), Au(111), Si(111), Si(557), Pb/Si, $NbSe_2$, HOPG samples with W, Au, Pt/Ir, Nb tips.

[4] *Protocol modification :* note that in some experiments we slightly modify the protocol: the tip was brought 2 Angstroms closer to the surface after switching the feedback loop off. The only found difference is in the enhanced reproducibility of the crater formation, but not in the aspect or in the size of the crater

**Figure 1. (a)** topographic STM image ($V_{bias}$ = 0.71 V, $I_T$ = 0.179 nA) acquired on a cleaved surface of $GaTa_4Se_8$ under Ultra High Vacuum. **(b)** Crystal structure of the (100) plane of $GaTa_4Se_8$ (Ta atoms: red, Se atoms: yellow, $GaSe_4$ tetraedra: green). The arrows indicate the crystallographic direction of the trenches.

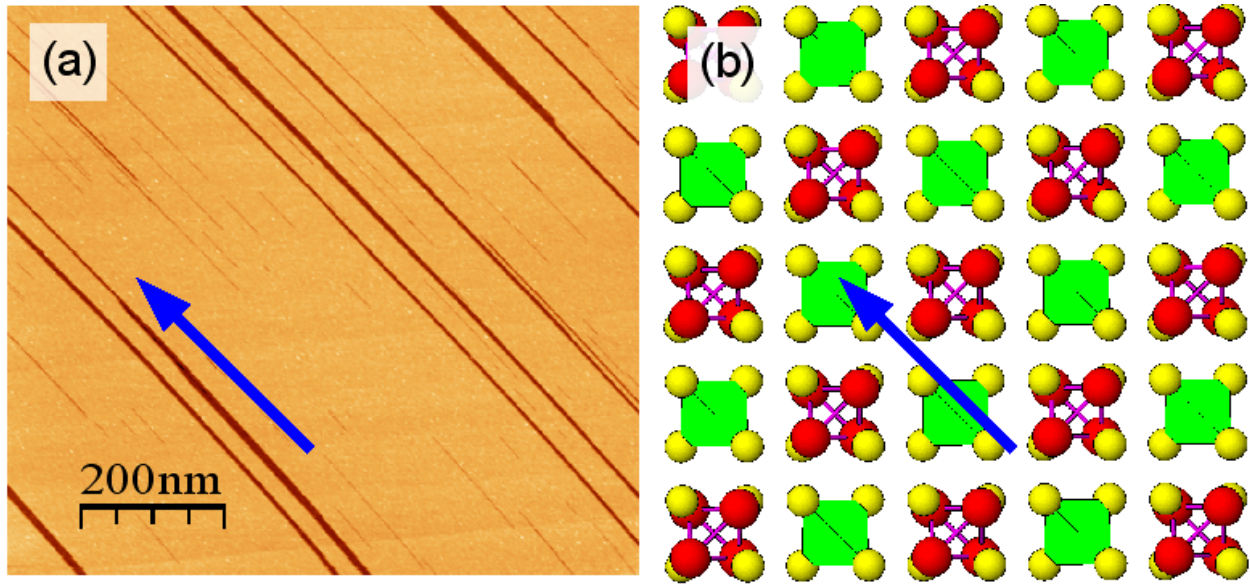

**Figure 2. (a)** Small scale topographic STM image ($V_{bias}$ = 0.80 V, $I_T$ = 0.11 nA) showing inhomogeneities with a characteristic size of ~3 nm. **(b)** The topographic profile corresponding to the black line in (a) showing the 1Å typical roughness of the surface.

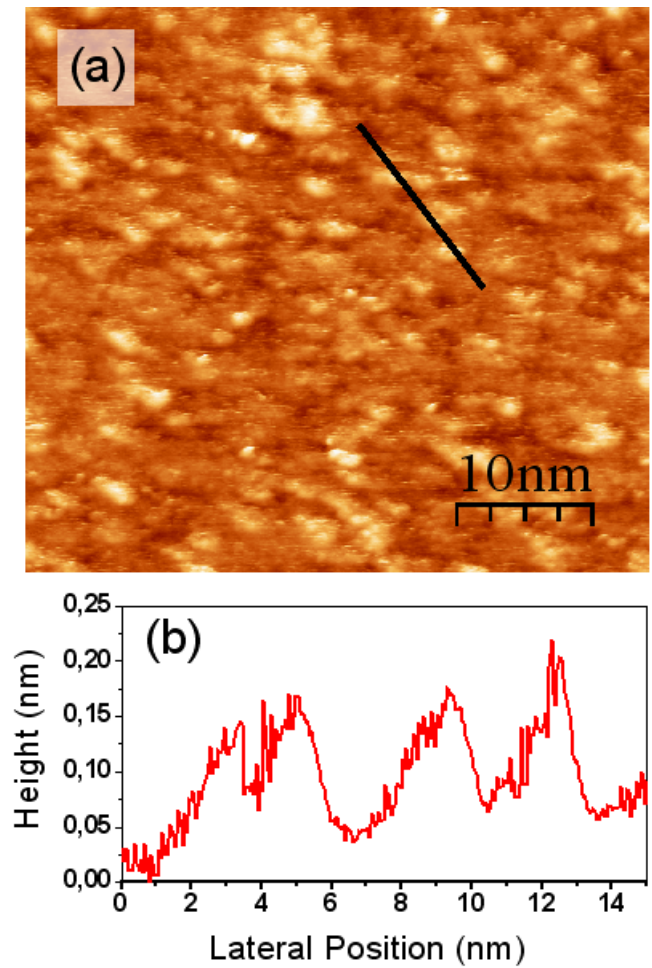

**Figure 3. (a)** Topographic STM image ($V_{bias}$ = 0.71 V, $I_T$ = 0.17 nA) showing the effect of duration and voltage of the pulses. Reference grid lines are made with 100ms@1.6V pulses. The test lines were done by varying the voltage pulse amplitude from 1.15 V to 1.9 V from left to right while the pulse duration varies from 100 µs to 100 ms from bottom to top. **(b)** Histogram showing the distribution of the crater radii of the reference grid. **(c)** Dependence of the crater radius as a function of the pulse voltage: Red dots represent the crater radius averaged with respect to the pulse duration, solid line is a linear fit.

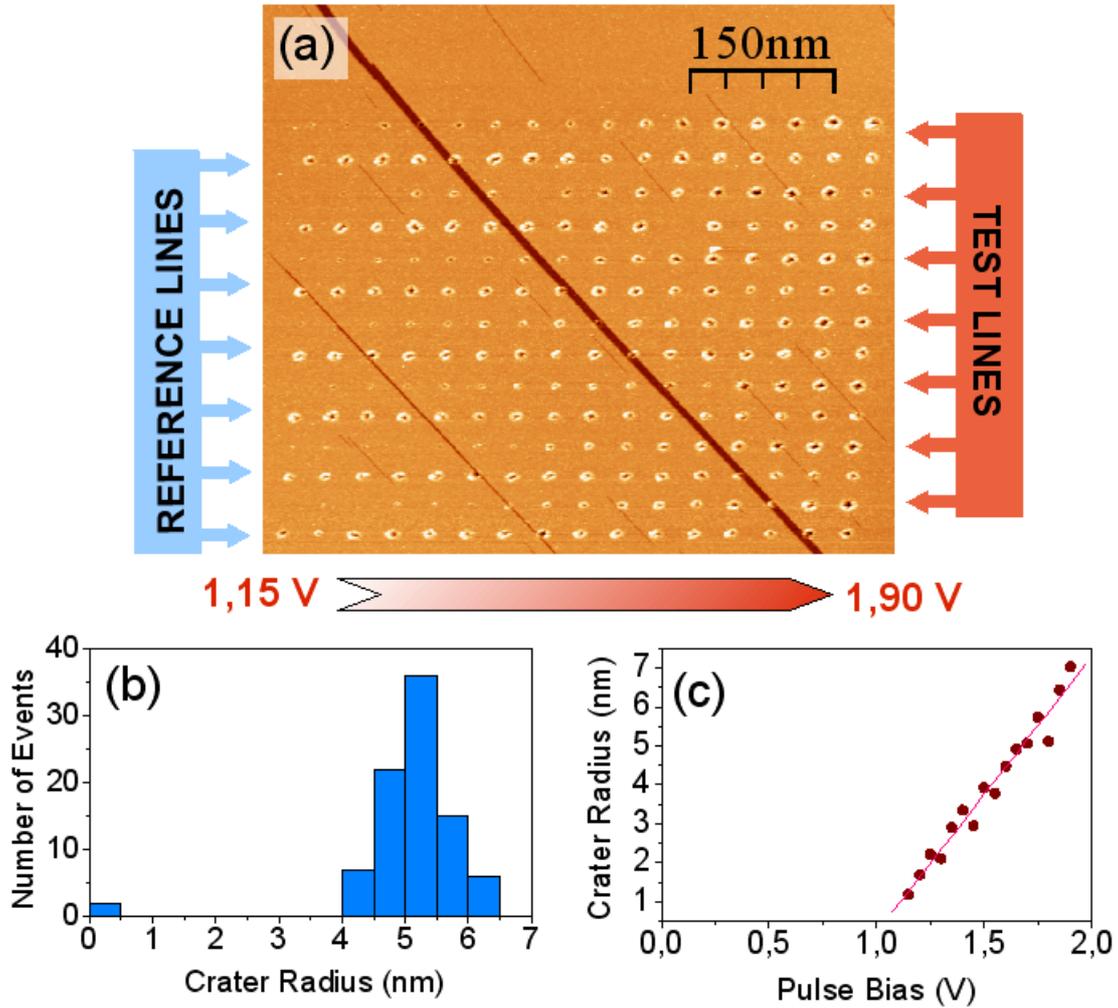

**Figure 4.** Topographic STM image ($V_{bias}$ = 0.758 V, $I_T$ = 0.186 nA) of a nanostructured pattern. The dots were done with 100 ms @ 1.6 V pulses. Note that a step of a half the unit-cell height is clearly visible at the top of the image.

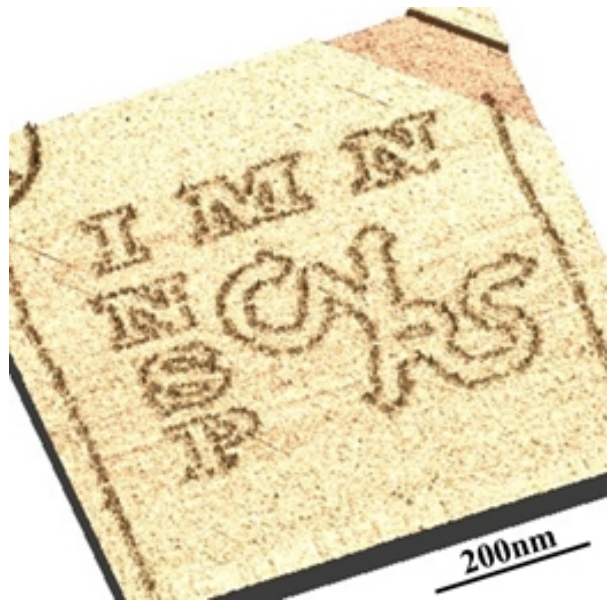

**The table of contents entry** We report the first experimental evidence for a strong electromechanical coupling in the Mott insulator $GaTa_4Se_8$ allowing a highly reproducible nano-writing with a Scanning Tunneling Microscope. Above a threshold voltage the surface inflates and comes in contact with the STM tip, resulting in nanometer size craters. These patterns can be indestructibly "read" by STM at lower voltage bias, thus allowing a 5 Tdots/inch$^2$ dense writing/reading at room temperature.

Keywords: Electric field effects, Mott insulator, Surface patterning, Scanning tunneling microscopy, Transition metal chalcogenides.

Vincent Dubost*, Tristan. Cren, Cristian Vaju, Laurent Cario, Benoit Corraze, Etienne Janod, François Debontridder, Dimitri Roditchev

Title : Electric Field Assisted Nanostructuring of a Mott Insulator

ToC figure ((55 mm broad, 50 mm high, or 110 mm broad, 20 mm high))

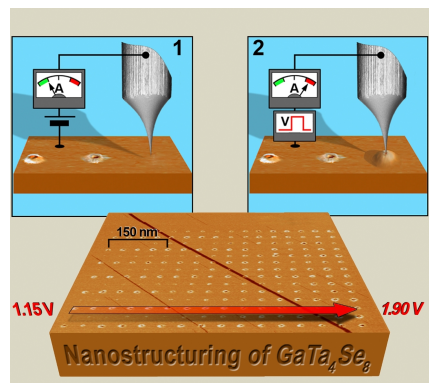